
\font\tenbf=cmbx10 scaled\magstep 1
\font\tenrm=cmr10
\font\tenit=cmti10
\font\elevenbf=cmbx10 scaled\magstep 1
\font\elevenrm=cmr10 scaled\magstep 1
\font\elevenit=cmti10 scaled\magstep 1

\rightline{MC/TH 94/16}
\hsize=6.0truein
\vsize=8.8truein
\parindent=3pc
\baselineskip=10pt
\vglue 12pt
\centerline{\tenbf NONPERTURBATIVE EFFECTS IN LOW ENERGY}
\vglue 0.2cm
\centerline{{\tenbf EFFECTIVE THEORIES OF QCD} \ \footnote\ddag{Talk
delivered at the Workshop on Quantum Infrared Physics, AUP, Paris,
 9th June 1994.}}
\vglue 1.0cm
\centerline{\tenrm DIMITRI KALAFATIS}
\baselineskip=13pt
\centerline{\tenit Theoretical Physics Group}
\baselineskip=12pt
\centerline{\tenit Department of Physics and Astronomy}
\baselineskip=12pt
\centerline{\tenit University of Manchester}
\baselineskip=12pt
\centerline{\tenit MANCHESTER M13 9PL, UK}
\vglue 0.8cm
\centerline{\tenrm ABSTRACT}
\vglue 0.3cm
{\rightskip=3pc
\leftskip=3pc
\tenrm\baselineskip=12pt
\noindent In this talk I concentrate on the role of chiral symmetry realisation
by spin-1 fields in the low energy QCD effective lagrangian. I assume that
chiral symmetry is nonlinearly realised and that spin-1 fields transform
homogeneously under chiral rotations, which is in sharp contrast to previous
works
where the $\rho$ and the $a_1$ mesons were treated as approximate gauge bosons
of some chiral group. I emphasise the role played by four-meson couplings which
in our scheme are essential for the theory to make sense. By requiring the
kinetic energy density of the theory to be bounded from below we find
inequalities relating three- and four-point meson couplings. It is finally
shown
how the combination of our analysis and of unitarity requirements naturally
leads to a low-energy phenomenological lagrangian for the nonanomalous sector
of
$\pi\rho a_1$ strong interactions.

\vglue 0.8cm }
\line{\elevenbf 1. Introduction \hfil}
\bigskip
\baselineskip=14pt
\elevenrm
Strong interactions at low energies are quite well understood in terms of an
effective
meson lagrangian$\displaystyle{^1}$. The starting point for such an effective
lagrangian is the nonlinear sigma model of pseudoscalar pions with
spontaneous breaking of chiral symmetry, a central feature of low energy QCD.
The experimental discovery of meson resonances as well as some theoretical
notions such as the large $N_c$ expansion of QCD$\displaystyle{^2}$, strongly
support
the idea of introducing mesons other than the pion into this model. There is
a considerable amount of work in the literature treating the role played by
massive spin-$1$ mesons (the $\displaystyle\rho$- and the
$\displaystyle{a_1}$-mesons) in
low-energy lagrangians. In most of these works isovector resonances are
introduced as
massive Yang-Mills particles$\displaystyle{^3}$ or as gauge bosons of local
chiral
symmetry$\displaystyle{^4}$, and low energy phenomena such as $\rho$-coupling
universality, the KSRF relation are then nicely described.

It should be made clear however that there is neither
experimental evidence nor theoretical prejudice from QCD to support an even
approximate
dynamical gauge boson character of the spin-1 resonances and therefore justify
among other things the conspicious emergence of Yang-Mills self-couplings
of the $\displaystyle{\rho}$ and the $\displaystyle{a_1}$ fields. As we stated
above,
the nice feature of the \lq\lq gauge" models is their natural compliance with
the phenomenologically successful notion of vector meson dominance, but as
other
authors have shown$\displaystyle{^5}$, this feature is not unique to the models
of
Refs. 3-4. It can also be obtained in cases where chiral symmetry is realised
in a less exotic manner.

In our approach we construct a lagrangian consistent with general
principles of quantum field theory and chiral symmetry. Vector
meson dominance can be implemented later, if so desired. We
assume therefore a homogeneous transformation law for isovector spin-1
fields. One of the purposes of my talk is to convince you that even in that
case constraints relating three- and four-point coupling strengths do exist.
These derive from demanding the hamiltonian to be bounded from below. I will
extend our previous analysis of the $\displaystyle{\pi\rho}$
system$\displaystyle{^6}$
to the description of interacting pions, $\rho$- and $a_1$-mesons.
In section 2 I introduce you to the symmetry structure of our scheme and build
the
basic interaction lagrangian. Section 3 is
devoted to the investigation of the energies of some nonperturbative field
configurations. It is shown that these energies are unbounded from below at the
three-meson coupling
level. In section 4 I demonstrate how the inclusion of four-point effective
couplings counterbalances the dangerous contributions of
the three-point terms to these energies and derive inequalities between three-
and four- meson couplings for the theory to make sense. We finally discuss in
section 5 how unitarity arguments based on vector dominance could lead to
saturation of
these inequalities and suggest a novel low-energy $\pi\rho a_1$ effective
lagrangian
consistent with chiral symmetry and general field theoretical principles.
\vglue 0.6cm
\line{\elevenbf 2. Nonlinear realisations of chiral symmetry \hfil}
\vglue 0.4cm

Let us start with the lagrangian of the nonlinear sigma model defined in
terms of the $SU(2)$ field $U$ as:
$$\eqalign{
{\cal{L}}_{NL\sigma}={{f^2}\over 4}<\partial^\mu U \partial_\mu
U^\dagger >,\cr}\eqno(1)$$
$f$ being the pion decay constant. We define $U$ as
$U=\exp{(i\vec\tau.\vec F(x))}$ with the pion field given by $\vec F=F \hat F$.
Other parametrisations are perhaps more suitable for perturbative evaluations
of Green' s functions, but are not as convenient for investigations of the
large field region, which is of interest for our purposes.

The lagrangian (1) is invariant under the linear
$SU(2)_L\otimes SU(2)_R$ group rotation $U\to g_L U g_R^\dagger$.
It is also invariant under the following nonlinear
rotation$\displaystyle{^7}$:
$$\eqalign{
u(\vec F) \ \to \ g_L u(\vec F) h^\dagger(\vec F) \ = \ h(\vec F)
u(\vec F) g_R^\dagger,\cr}
\eqno(2)
$$
where $u$ is the square root of $U$ and $h(\vec F)$ is a compensating
transformation ensuring that $U$ transforms linearly. Consider now
the following axial-vector and vector respectively field gradients:
$$\eqalign{
u_\mu=&i (u^\dagger\partial_\mu u-u\partial_\mu u^\dagger)\cr
\Gamma_\mu=&{1\over 2}(u^\dagger\partial_\mu u+u\partial_\mu u^\dagger).\cr}
\eqno(3)
$$
These gradients transform as
$$\eqalign{
u_\mu \ \to& \ h(\vec F) u_\mu h^\dagger (\vec F) \cr
\Gamma_\mu \ \to& \ h(\vec F) \Gamma_\mu h^\dagger (\vec F) +
h(\vec F) \partial_\mu h^\dagger (\vec F).\cr}\eqno(4)
$$
It is seen that while $u_\mu$ transforms
homogeneously the transformation of
$\Gamma_\mu$ contains an inhomogeneous part as a result of
the field dependence of $h(\vec F)$. How should spin-1 fields
transform in this framework? Basically there are two
possibilities forming a group:homogeneous or inhomogeneous. The
inhomogeneous group was chosen by Weinberg$\displaystyle{^1}$ in its study
of the $\pi\rho$ system. This approach later acquired the name
of \lq\lq hidden gauge symmetry"$\displaystyle{^4}$ approach,
because the associated lowest order invariant lagrangian
preserves not only chiral symmetry but also a
certain sort of a gauge symmetry. It is one of my purposes to point out that
this symmetry was never revealed by low energy $\pi\pi$ scattering
experiments. Furthermore in the inhomogeneous approach it is impossible to
define
similar transformations for both the $\rho$ and the $a_1$ fields, simply
because the associated particles have opposite parity.

In contrast the homogeneous transformation is the simplest one consistent
with chiral symmetry and it can be applied to both the $\rho$ and the $a_1$
fields. We
adopt this point of view and assume that the $\rho$- and the
$a_1$-mesons transform as
$$\eqalign{
V_\mu \ \to &\ h(\vec F) V_\mu h^\dagger(\vec F)\cr
A_\mu \ \to &\ h(\vec F) A_\mu h^\dagger(\vec F)\cr}\eqno(5)
$$
under the nonlinear group. In this expression $V_\mu=\vec\tau.\vec V_\mu$
and $A_\mu=\vec\tau.\vec A_\mu$. To build invariant couplings we define
covariant derivatives of spin-1 fields transforming as in eq.~(5)
$$\eqalign{
\nabla_{\mu} =\partial_{\mu} +
[\Gamma_\mu, \ ],\cr}\eqno(6)
$$
in such a way that $\nabla_\mu V_\nu$ and $\nabla_\mu
A_\nu$ also transform homogeneously: $\nabla_\mu V_\nu\to
h\nabla_\mu V_\nu\ h^\dagger$ and similarly $\nabla_\mu A_\nu\to
h \nabla_\mu A_\nu\ h^\dagger$.

\noindent The invariant lagrangian at quadratic order in the fields is given by
$$\eqalign{
{\cal{L}}_{\pi\rho a_1}^{(2)}=&{{f^2}\over 4}<u_\mu u^\mu>-
{1\over 4}<V_{\mu\nu}V^{\mu\nu}>
-{1\over 4}<A_{\mu\nu}A^{\mu\nu}>\cr
&+{{M_{\rho}^2}\over 2}<V_\mu V^\mu>+
{{M_a^2}\over 2}<A_\mu A^\mu>,\cr}\eqno(7)
$$
where $V_{\mu\nu}=\nabla_\mu V_\nu-\nabla_\nu V_\mu$ and $A_{\mu\nu}=\nabla_\mu
A_\nu-\nabla_\nu A_\mu$ are the covariant field strengths of the spin-1
resonances. We introduce chirally invariant mass terms for the $\rho$- and the
$a_1$-mesons and we assume that the coupling $c< A_\mu u^\mu >$ is not present.
Actually with the choice $c=0$ no diagonalisation of $\pi\rho a_1$
interactions is needed - obviously not a disadvantage of our framework.
At the three-point level we shall consider some chirally invariant terms
consistent with charge conjugation and parity invariance, containing
at least one pion field gradient:
$$\eqalign{
{\cal{L}}_{\pi\rho a_1}^{(3)}=-{i\over{2\sqrt{2}}}\bigg\{ \ &g_1 <V_{\mu\nu}
[u^\mu,u^\nu]>  \ +  \ g_2 <A_{\mu\nu}\big([V^\mu,u^\nu]-[V^\nu,u^\mu]\big)>\cr
+ &g_3 <V_{\mu\nu} \big([A^\mu,u^\nu]-[A^\nu,u^\mu]\big)> \
\bigg\}.\cr}\eqno(8)
$$
The lagrangian ${\cal{L}}_{\pi\rho a_1}^{(2)}+{\cal{L}}_{\pi\rho a_1}^{(3)}$
has six free parameters that one would ideally like to determine
from QCD. But this problem seems to be elusive since a sensible method to
perform such an extraction from QCD has not yet been discovered.

\vglue 0.6cm
\line{\elevenbf 3. Nonperturbative pathologies\hfil}
\vglue 0.4cm

The issue I address here is rather different: assuming that $g_1, g_2, g_3$
are given by the underlying QCD dynamics, are there any relations
between these parameters and higher order ones? Previous
investigations$\displaystyle{^6}$
suggest that this question should be addressed in a nonperturbative framework.
In
particular does the theory ~(7, 8) yield a hamiltonian that is bounded from
below?
To find an answer let us investigate the hamiltonian associated with the
lagrangian
${\cal{L}}_{\pi\rho a_1}^{(2)}+{\cal{L}}_{\pi\rho a_1}^{(3)}$ in terms of the
canonical degrees of freedom: the fields ${\vec F,\ \vec V_i,\ \vec A_i}$ and
their conjugate momenta, respectively ${\vec \phi,\ \vec \pi_i, \vec \chi_i}$.
The energy can be written as a sum of two terms $H=H_T+H_V$,
where the kinetic energy is $H_T$ and the potential energy is $H_V$. The
potential part contains only space components and in the three-point
case is given by
$$\eqalign{
H_V=&\int d^3 x \bigg\{
{{f^2}\over 2}(u_i)_k^2+M_a^2(A_i)_k^2
+{1\over 2}(A_{ij})_k\big[ \ A_{ij}+i\sqrt{2}g_2([V_i,u_j]-[V_j,u_i])\
\big]_k\cr
&+M_{\rho}^2(V_i)_k^2+{1\over 2}(V_{ij})_k\big[ \ V_{ij}+i\sqrt{2}g_1[u_i,u_j]
+i\sqrt{2}g_3([A_i,u_j]-[A_j,u_i]) \ \big]_k\bigg\}.\cr}\eqno(9)
$$
The kinetic piece needs some work in order to eliminate the dependent variables
$\vec V_0,\ \vec A_0$. It turns out that the generic expression for $H_T$ in
the case
of a quadratic in momenta theory is:
$$\eqalign{
H_T=\int d^3 x \bigg\{
{1\over 2} \vec\Phi   {\cal{A}}^{-1} \vec\Phi
+{{\vec\pi_i^2}\over 4} + {{\vec\chi_i^2}\over 4}
+{1\over 2} \vec\Gamma {\cal{P}}^{-1} \vec\Gamma \bigg\},\cr}\eqno(10)
$$
where $\vec\Phi,\ \vec\Gamma$ are linearly related to the momenta
${\vec \phi,\ \vec \pi_i,\ \vec \chi_i}$ and ${\cal{A}},\ {\cal{P}}$ are
isospin tensor functions of ${\vec F,\ \vec V_i,\ \vec A_i}$, which depend on
the detailed structure of the dynamics. Let us now investigate the energy
of a classical meson mapping. The simplest such object one can imagine has
an isospin content specified by a constant unit vector $\hat F$:
$$\vec F_0(\vec x)= F(\vec x)\ \hat F, \eqno(11)$$
where $F(\vec x)$ is a regular function of space. Such a configuration is
topologically trivial. For the vector and the axial vector fields I will assume
that they are parallel to the pion field:
$$\eqalign{
\vec V_i(\vec x)=& V_i(\vec x) \ \hat F \cr
\vec A_i(\vec x)=& A_i(\vec x) \ \hat F .\cr}\eqno(12)
$$
The special form of our ansatz result in a potential energy which is simply
given
by:
$$\eqalign{
H_V=\int d^3 x \bigg\{
{{f^2}\over 2}(\partial_i F)^2+M_a^2A_i^2
+{1\over 2}(\partial_{i}A_{j}-\partial_j A_i)^2+M_{\rho}^2V_i^2+
{1\over 2}(\partial_{i}V_{j}-\partial_j V
_i)^2\bigg\},\cr}\eqno(13)
$$
the potential energy of a free theory, an obviously positive quantity.
We therefore concentrate in the kinetic energy of the theory as given
by $H_T$. Because of the particular isospin structure we
consider here only momenta that point in a direction perpendicular
to that of the pion actually \lq\lq see" the couplings to the
vector mesons. We assume:
$$\eqalign{
\vec \phi=& \phi(\vec x) \ \hat \phi\cr
\vec \pi_i=& \pi_i(\vec x) \ \hat F \cr
\vec \chi_i=& \chi_i(\vec x) \ \hat F,\cr}\eqno(14)
$$
with $\hat\phi\cdot\hat F=0$. The kinetic energy of this field
configuration reads
$$\eqalign{
H_T=\int d^3 x \bigg\{{{\phi^2}\over{2f^2s^2 {\cal{I}}}}
+{1\over 4}\bigg[\pi_i^2+\chi_i^2+{{(\partial_i\pi_i)^2}\over{M_\rho^2}}
+{{(\partial_i\chi_i)^2}\over{M_a^2}}\bigg]\bigg\},\cr}\eqno(15)
$$
where $s$ is a shorthand notation for $\sin{F}/F$. Observe now
the structure of the dimensionless \lq\lq inertial" parameter ${\cal{I}}$
which contains all the nontrivial effects due to the spin-1 fields:
$$\eqalign{
{\cal{I}}={1\over{f^4M_1M_2}}\bigg(&f^4M_1M_2-8(g_2^2 V_i^2M_2M_\rho^2
+(g_1\partial_i F-g_3 A_i)^2M_1M_a^2)\cr
&+16\big[g_2^4M_2(V_i^2(\partial_i F)^2-(V_i\partial_i F)^2)
+g_3^4M_1(A_i^2(\partial_i F)^2-(A_i\partial_i F)^2)\big]\bigg),\cr}\eqno(16)
$$
with $M_1=(1/f^2)\displaystyle{\big[2M_\rho^2-4g_2^2(\partial_i F)^2\big]}$ and
$M_2=(1/f^2)\displaystyle{\big[2M_a^2-4g_3^2(\partial_i F)^2\big]}$.
While in the case of the nonlinear sigma model with vanishing
couplings ${\cal{I}}$ is simply equal to $1$, here it acquires
negative contributions. For small fields $F,\ V_i,\ A_i\approx 0$
appropriate to perturbation theory one has ${\cal{I}}\approx 1$
so no problem seems to appear in perturbative expansions of scattering
amplitudes. For nonperturbative configurations however
the situation changes dramatically since then negative contributions
proportional to quadratic powers of the couplings can drive ${\cal{I}}$
to zero or negative values. The hamiltonian density acquires poles and
the energy is not bounded from below. Such troubles can arise at
fairly low energy scales. Consider a localised meson wave
carrying momentum $k_i$ and of amplitude $F \approx 1$. Assume to simplify
further that all classical fields
vanish except $F$ and $\phi$. The gradient $\partial_i F$ is roughly
approximated by $k_i$ and ${\cal{I}}$ inside the meson wave looks like:
$$\eqalign{
{\cal{I}}\approx \displaystyle{{1-2\bigg(\displaystyle{{g_3^2}\over{M_a^2}}
+2\displaystyle{{g_1^2}\over{f^2}}\bigg)k^2}\over{
1-2\displaystyle{{g_3^2}\over{M_a^2}}k^2}}.\cr}\eqno(17)
$$
At small or very large momenta $k$ the inertial parameter is positive but for
$k^2$ in the intermediate range
$$\eqalign{
\left(2{g_3^2\over M_a^2}+4{g_1^2\over f^2}\right)^{-1} < k^2 <
{M_a^2\over 2g_3^2}\cr}\eqno(18)
$$
${\cal{I}}$ becomes negative and as a consequence the kinetic energy
density is negative, making the theory ill defined in these regions. Taking
reasonable
numerical values for the coupling constants$\displaystyle{^8}$, the region
of dangerous momenta is found to be $0.4$ GeV $< k < 2.0$ GeV, which
includes the range of masses of the $\rho$ and the $a_1$ resonances. And
this is precisely the range that one would like to describe by extending the
low-energy effective theories to include spin-1 mesons !

We conclude that the hamiltonian associated with the simplest three-point
$\pi\rho a_1$ interactions {\elevenit is not bounded from below} which is of
course
unacceptable.
\vglue 0.6cm
\line{\elevenbf 4. Constraints on four-meson coupling strengths \hfil}
\vglue 0.4cm

Let us now consider the effect of some four-meson couplings that are
relevant to our analysis:
$$\eqalign{
{\cal{L}}_{\pi\rho a_1}^{(4)}=&{1\over 8}\bigg\{g_4<[u_\mu,u_\nu]^2>+2g_5
<[u_\mu,u_\nu][A^\mu,u^\nu]>+2g_6\big(<[V_\mu,u_\nu]^2>\cr
&-<[V_\mu,u_\nu][V^\nu,u^\mu]>\big)+2g_7\big(<[A_\mu,u_\nu]^2>-<[A_\mu,u_\nu]
[A^\nu,u^\mu]>\big)\bigg\}.\cr}\eqno(19)
$$
We introduce four new coupling constants $g_4,\ g_5,\ g_6,\ g_7$.
Amongst these terms one can recognise a local four-point pion vertex, the
so-called \lq\lq Skyrme term", as well as a term contributing to the decay
$a_1\to \pi\pi\pi$.

Concerning now the energy of the charge-zero meson configuration defined in the
previous section, we note first that its potential energy is unaffected by the
new couplings and is still given by eq.~(13). The kinetic piece has the same
form as in eq.~(15) but with a new inertial function, $\tilde{\cal{I}}$. After
a tedious but straightforward calculation one finds:
$$\eqalignno{
\tilde{\cal{I}}={1\over{f^4\tilde{M}_1\tilde{M}_2}}
\bigg\{&f^4\tilde{M}_1\tilde{M}_2-8(g_2^2-g_6)V_i^2M_\rho^2\tilde{M}_2\cr
&+16\bigg[(g_2^2-g_6)^2\big(V_i^2(\partial_i F)^2
-(\partial_i F V_i)^2\big)\bigg]\tilde{M}_2\cr
&-8\bigg[(g_1^2-g_4)(\partial_i F)^2+(g_3^2-g_7)A_i^2-2\left(g_3g_1-{{g_5}
\over 2}\right)(\partial_i F A_i)\bigg]M_a^2 \tilde{M}_1\cr
&+16 \bigg[\bigg((g_1^2-g_4)(g_3^2-g_7)-\bigg(g_3g_1-{{g_5}\over 2}\bigg)^2
\bigg)(\partial_i F)^4&(20)\cr
&\qquad+(g_3^2-g_7)^2\big(A_i^2(\partial_i F)^2
-(A_i\partial_i F)^2\big) \bigg]\tilde{M}_1\bigg\},\cr}
$$
with $\tilde{M}_1=(1/f^2)\displaystyle{\big[2M_\rho^2-4(g_2^2-g_6)(\partial_i
F)^2\big]}$ and
$\tilde{M}_2=(1/f^2)\displaystyle{\big[2M_a^2-4(g_3^2-g_7)(\partial_i
F)^2\big]}$. By requiring that for any value of the classical profiles
$\partial_i F,\
V_i,\ A_i$ the function $\tilde{\cal{I}}$ is non-negative, we find constraints
{\elevenit
on the couplings}. Consider the following simplifying cases:

\medskip

\noindent {\hskip 0.5cm} $\displaystyle{ a) \ \ \ \partial_i F=A_i=0 \ \ \ \ \
\Rightarrow \ \ \ \ \ \tilde{\cal{I}}_a=1-{4\over{f^2}}(g_2^2-g_6)V_i^2}$

\bigskip

\noindent {\hskip 0.9cm} $\displaystyle{ b) \ \ \ \partial_i F=V_i=0 \ \ \ \ \
\Rightarrow \ \ \ \ \ \tilde{\cal{I}}_b=1-{4\over{f^2}}(g_3^2-g_7)A_i^2}$
{\hfill} $(21)$
$$\eqalign{
c) \ \ \ V_i=A_i=0 \ \ \ \ \ \ \ \Rightarrow \ \ \ \ \ \tilde{\cal{I}}_c
=&\displaystyle{1\over{f^2\tilde{M}_2}}\bigg[2M_a^2-\big[4(g_3^2-g_7)
+8(g_1^2-g_4)\displaystyle{{M_a^2}\over{f^2}}\big](\partial_i F)^2\cr
&+\displaystyle{{16}\over{f^2}}\big[(g_1^2-g_4)(g_3^2-g_7)-(g_3g_1-{{g_5}
\over 2})^2\big](\partial_i F)^4\bigg].\cr}
$$
Requiring positiveness of $\tilde{\cal{I}}_{a,b,c}$ for all possible values
of the fields leads to:
$$\eqalign{
g_4 \  \ge& \ g_1^2\cr
g_6 \ \ge& \ g^2_2\cr
g_7 \  \ge& \ g_3^2\cr
(g_1^2-g_4)(g_3^2-g_7) \ \ge& \ (g_3g_1-{{g_5}\over 2})^2.\cr
}\eqno(22)
$$

These conditions show that the Skyrme term and other four-point
interactions are essential if the hamiltonian is to be bounded from below. In
an effective theory where the spin-1 fields transform homogeneously they arise
as counterterms for the bad behaviour of the vector-meson
contributions. This is in sharp contrast to other
approaches$\displaystyle{^{3,4}}$, where the same Skyrme term
emerges from the exchange of a very heavy $\rho$-meson.

\vglue 0.6cm
\line{\elevenbf 5. Discussion \hfil}
\vglue 0.4cm

Our investigation of classical nonperturbative effects in low-energy chiral
theories shows that the constraints (22), relating
three- and four-point couplings, must be satisfied for a consistent description
of the interactions between pions and spin-1 isovector mesons. We stress that
chiral symmetry is implemented nonlinearly in this approach and the vector
mesons are naturally assumed to transform homogeneously under chiral rotations.
The constraints arise from demanding the hamiltonian to be bounded from
below. They do not depend on phenomenological ideas such as vector dominance.

One might ask now whether there are any other constraints on the couplings from
first principles. For instance another nonperturbative notion that one could
invoke in this context is the unitarity of the scattering matrix. This
was previously studied$\displaystyle{^{9}}$ in the special case of
the lagrangian (7, 8) without the $a_1$ ($g_2=g_3=0$). Working at
tree-level it was found that further local pion interactions must be
added by hand if the forward elastic $\pi\pi$ scattering amplitude is to obey
the
Froissart bound$\displaystyle{^{10}}$. These local interactions compensate for
the most divergent contribution produced by $\rho$-exchange.
The result, in the SU(2) sector is:
$$\eqalign{
{\cal{L}}^{SU(2)}_{local}={{g_1^2}\over 8}  < [u_\mu,u_\nu]^2 >.\cr}\eqno(23)
$$
This is just the Skyrme term, with a coefficient that is fixed by the
three-point coupling $g_1$. If one works at tree level, the $a_1$ does not
contribute
to $\pi\pi$ scattering. Imposing unitarity therefore leads to saturation
of the lower bound on $g_4$ in (22):
$$\eqalign{
g_4-g_1^2=0.\cr}\eqno(24)
$$
Combining this with the final constraint in (22), we obtain a relation
expressing the implications  of unitarity for the couplings of the {\elevenit
axial}
meson:
$$\eqalign{
g_5=2g_1g_3.\cr}\eqno(25)
$$
This is nontrivially relating the strength of the $a_1\to\pi\pi\pi$
decay to those of the processes $\rho\to\pi\pi$ and $a_1\to\rho\pi$.

The saturation of two of our constraints follows from the assumption that
an extreme version of vector dominance holds for strong interactions.
In fact the value for $g_4$ determined assuming $\rho$-meson dominance
agrees well with that from chiral perturbation theory$\displaystyle{^{11}}$,
suggesting that at least at low energies, vector dominance is really making
sense.
We speculate that dominance of a single resonance may also hold in the
axial-vector
channel, leading to saturation of the remaining constraints in (24).
In this case our lagrangian would simplify to:
$$\eqalign{
{\cal{L}}_{\pi\rho a_1}=&{{f^2}\over 4}<u_\mu u^\mu>+{{M_{\rho}^2}\over
2}<V_\mu V^\mu>+
{{M_a^2}\over 2}<A_\mu A^\mu>\cr
-&{1\over
4}<\bigg(V_{\mu\nu}+{i\over{\sqrt{2}}}\big(g_1[u_\mu,u_\nu]+g_3([A_\mu,u_\nu]-[A_\nu,u_\mu])\big)\bigg)^2>\cr
-&{1\over 4}<\bigg(A_{\mu\nu}
+{i\over{\sqrt{2}}}g_2([V_\mu,u_\nu]-[V_\nu,u_\mu])\bigg)^2>.\cr}\eqno(26)
$$
This constitutes an effective lagrangian describing the strong interactions of
$\pi\rho a_1$ mesons with a minimal number of free coupling constants. It is
the simplest one compatible with chiral symmetry and leading to a hamiltonian
which is free of pathologies.

To summarise: in our framework theory chiral symmetry is
implemented in the simplest possible way and no speculative gauge symmetry
assumption is made. Constraints between couplings are there to ensure that the
hamiltonian is bounded from below, and vector meson dominance can be
implemented by specific choices of parameters. It is therefore most natural
to regard our lagrangian (26) as the starting point for any extension of chiral
perturbation theory of pseudoscalar pions to the resonance region.

\vglue 0.6cm
\line{\elevenbf 6. Acknowledgements \hfil}
\vglue 0.4cm
This work was done in collaboration with Mike Birse. I
would like to thank him for
his contribution to my understanding of low energy phenomena.
I gratefully acknowledge financial support from the SERC.
\vglue 0.6cm
\line{\elevenbf 7. References \hfil}
\vglue 0.4cm
\item{1.}S. Weinberg, {\elevenit Phys. Rev.} {\elevenbf 166} (1968) 1568;
I. Gerstein, R. Jackiw, B. Lee and S. Weinberg, {\elevenit Phys. Rev.}
{\elevenbf D3} (1971) 2486;
A. Salam and J. Strathdee, {\elevenit Phys. Rev.} {\elevenbf D2} (1970) 2869
\item{2.}G. 't Hooft, {\elevenit Nucl. Phys.} {\elevenbf B79} (1974) 276
\item{3.}J. J. Sakurai, {\elevenit Currents and Mesons}, University of Chicago
Press, Chicago 1969;
J. Schwinger, {\elevenit Particles and Sources}, Clarendon Press, Oxford, UK,
1969
\item{4.}M. Bando, T. Kugo and K. Yamawaki, {\elevenit Phys. Rep.} {\elevenbf
164} (1988) 217
\item{5.}G. Ecker, J. Gasser, A. Pich and E. de Rafael, {\elevenit Nucl. Phys.}
{\elevenbf B321} (1989) 311
\item{6.}D. Kalafatis, {\elevenit Phys. Lett.} {\elevenbf B313} (1993) 115
\item{7.}S. Coleman, J. Wess and B. Zumino, {\elevenit Phys. Rev.} {\elevenbf
177} (1969) 2239;
C. G. Callan, S. Coleman, J. Wess and B. Zumino, {\elevenit ibid} 2247
\item{8.}B. Moussallam, private communication
\item{9.}G. Ecker, J. Gasser, H. Leutwyler, A. Pich and E. de Rafael,
{\elevenit Phys. Lett.}  {\elevenbf B223} (1989) 425
\item{10.}M. Froissart, {\elevenit Phys. Rev.} {\elevenbf 123} (1961) 1053; A.
Martin, {\elevenit Phys. Rev.} {\elevenbf 129}
 (1963) 1432
\item{11.}J. Gasser and H. Leutwyler, {\elevenit Ann. Phys.} (NY) {\elevenbf
158} (1984) 142
\bye